\documentclass[aps,pre,reprint,groupedaddress,showpacs]{revtex4-1}

\usepackage{graphicx} 
\usepackage{comment}
\usepackage{color}
\usepackage{ulem}

\begin{document}


\title{Robustness of cooperation on scale-free networks
under continuous topological change}


\author{Genki Ichinose}
\email[]{ichinose@anan-nct.ac.jp}
\homepage[]{https://sites.google.com/site/igenki/}
\altaffiliation{}
\affiliation{Department of Systems and Control Engineering, Anan National College of Technology, 265 Aoki Minobayashi,
Anan, Tokushima 774-0017, Japan}

\author{Yuto Tenguishi}
\affiliation{Department of Systems and Control Engineering, Anan National College of Technology, 265 Aoki Minobayashi,
Anan, Tokushima 774-0017, Japan}

\author{Toshihiro Tanizawa}
\altaffiliation{}
\affiliation{Department of Electrical Engineering and Information Science, Kochi National College of Technology, 200-1 Monobe-Otsu,
Nankoku, Kochi 783-8508 Japan}


\date{\today}

\begin{abstract}
 In this paper, we numerically investigate the robustness of cooperation
 clusters in prisoner's dilemma played on scale-free networks,
 where the network topologies change
 by continuous removal and addition of nodes.
 Each removal and addition can be either random or intentional.
 We therefore have four different strategies in changing network topology:
 random removal and random addition (RR),
 random removal and preferential addition (RP),
 targeted removal and random addition (TR),
 and targeted removal and preferential addition (TP).
 We find that cooperation clusters are most fragile against TR,
 while they are most robust against RP,
 even for large values of the temptation coefficient for defection.
 The effect of the degree mixing pattern
 of the network is not the primary factor
 for the robustness of cooperation under
 continuous change in network topology,
 which is quite different
 from the cases observed in static networks.
 Cooperation clusters become more robust
 as the number of links of hubs occupied by cooperators increase.
 Our results might infer the fact that a huge variety of individuals is needed
 for maintaining global cooperation in social networks in the real world
 where each node representing an individual is constantly removed and added.
\end{abstract}

\pacs{89.75.Fb, 02.50.Le, 87.23.Kg, 87.23.Ge}

\maketitle


 \section{Introduction}
 
 The emergence of cooperation is one of the challenging problems
 in both social and biological sciences.
 Cooperators benefit others by incurring some costs to the actor
 while defectors do not pay any costs.
 Thus, under a well-mixed population, cooperation cannot evolve
 because defectors are always better off than cooperators. 
 This relationship between cooperators and defectors is well parametrized
 by the prisoner's dilemma (PD) game \cite{Axelrod1984}.
 In PD, two individuals decide whether to cooperate or defect simultaneously.
 They both obtain $R$ for mutual cooperation and $P$ for mutual defection.
 If one selects cooperation and the other selects defection,
 the former gets $S$ for being the sucker of the defector,
 and the latter gets $T$ as a reward for the temptation to defect.
 The order of the four payoffs is $T > R > P > S$ in PD.
 Nowak and May revealed that spatial structures are required
 for the evolution of cooperation \cite{NowakMay1992}.
 Recently, it has been possible to map any given spatial structure on a suitable network topology
 and the evolution of cooperation has been investigated
 through the analysis of PD played
 on the corresponding complex network
 \cite{AbramsonKuperman2001, MasudaAihara2003, Masuda2007,
 Ohtsuki_etal2006, Ashlock2007, SantosPacheco2005, Rong_etal2007,
 ZimmermannEguiluz2005, Tanimoto2009, SuzukiKatoArita2008,
 IchinoseKobayashi2011}.
 
 In this context, the spatial structure
 required for the emergence of cooperation is referred
 to as network reciprocity and becomes one of the most important factors
 for the emergence of cooperation \cite{Nowak2006}.
 If the network reciprocity and some other mechanisms are combined,
 cooperation is promoted more \cite{PercSzolnoki2010}.
 For instance, teaching activity \cite{SzolnokiPerc2008},
 social diversity \cite{PercSzolnoki2008},
 an ability to infer the reputation of others \cite{Wang_etal2012a},
 a sparse environment \cite{Wang_etal2012b, Wang_etal2012c},
 selecting high fitness individuals in adopting strategies
 \cite{WangPerc2010, PercWang2010},
 age structure \cite{Wang_etal2012d},
 and incorporating environmental factors as the fitness of the focal individual
 \cite{Wang_etal2011} promote cooperation when they are combined
 with the network reciprocity. 
 In these studies,
 various ``heterogeneities'' enhancing cooperation
 combined with network reciprocity are considered.
 On the other hand, networks can generate structural heterogeneity
 by themselves, which we refer to here as ``network heterogeneity.''
 For scale-free networks, this network heterogeneity
 is the key for promoting cooperation \cite{SantosPacheco2005}.
 If cooperators occupy hubs in a scale-free network surrounded by other cooperators,
 the payoffs for these cooperators are considerably
 higher than for other individuals.
 Thus, they can spread their cooperative strategy.
 In contrast, if defectors occupy hubs surrounded by other defectors,
 this cluster of defectors is quite vulnerable
 and is easily replaced by cooperators.
 These two effects contribute to the evolution of cooperation
 on scale-free networks.
 
 For scale-free networks,
 the robustness of cooperation has been examined
 on growing \cite{Poncela_etal2008} or
 reducing \cite{Perc2009} networks.
 Poncela {\it et al.}\ \cite{Poncela_etal2008} have proposed
 an evolutionary preferential attachment,
 in which high payoff nodes attract more links from new nodes.
 They introduced a control parameter,
 $\epsilon \in [0,1)$, for the preference weight.
 In the limit, $\epsilon \rightarrow 1$, each existing node can get a new link
 in proportion to its payoff.
 If $\epsilon = 0$, a new link is connected
 to any node at random,
 which means the payoffs are completely ignored.
 They have shown that cooperation is most promoted at the limit,
 $\epsilon \rightarrow 1$, resulting in a scale-free network
 because a center cooperator in the cooperative group
 has a high preference weight and tends to
 get a new link more easily.
 Moreover, this causes positive feedback of the increment of the degree
 of the center cooperator.
 The rich get richer.
 It should be noted, however, that
 the network growth in this model is only in the direction of increasing
 the number of nodes and that the opposite possibility of decreasing the number of nodes is totally ignored.
 
 On the other hand, Perc \cite{Perc2009} has studied the evolution of cooperation
 in the direction of decreasing the number of nodes.
 He has implemented two ways of node removal from the Bar\'{a}basi and Albert (BA) network model
 \cite{BarabasiAlbert1999}.
 One is random removal of a fraction of nodes $\Lambda$,
 and the other is targeted removal of nodes
 from the largest degree up to a fraction $\Lambda$.
 He has shown that the cooperation on scale-free networks
 is extremely robust against random node removal,
 while it declines rapidly against targeted attack.
 Notice that, in his model, removed nodes are never restored \cite{Perc2009}.
 However, in artificial networks, there are many cases
 in which the restoration of removed nodes immediately occurs.
 Likewise, in ecological networks,
 a vacant site due to the death of an individual is
 often filled with a new individual immediately.
 Therefore, it is plausible that a node removal is followed
 by an addition of another new node.
 The present paper deals with
 such a bidirectional network topological change
 and investigates the effects of continuous removal and addition of nodes
 in the evolution of cooperation.

 One of the other factors that potentially affects
 the robustness of cooperation is the degree correlation between nodes
 represented by degree mixing patterns, which was investigated by
 Rong {\it et al.}\ \cite{Rong_etal2007}.
 In their model, a network is referred to as assortative (disassortative)
 according to the tendency of highly connected nodes (hubs) to choose nodes
 with similar (dissimilar) degrees as neighbors.
 Rong {\it et al.} have shown that the assortative network favors defection
 because the hubs tend to connect closely, which allows
 defectors to invade cooperators.
 In contrast, cooperation is maintained in the disassortative networks
 because the isolation of hubs due to disassortativity
 enables them to keep their initial strategy.
 At the same time, however,
 the influence of the hubs becomes weaker as the disassortativity
 increases because the tendency of the isolation also increases.
 Therefore, uncorrelated networks promote cooperation
 to the maximum extent
 by spreading the strategy of hubs most effectively.
 This conclusion, nevertheless, only applies to networks with a static topology.
 Once we allow the change of network topology by removal and addition
 of nodes or links,
 the mixing patterns change accordingly,
 and the conclusion observed in static networks might fail to apply.
 It is therefore also worth investigating
 the effects of the alteration of degree correlation caused
 by continuous node removal and addition
 on the evolution of cooperation.

 In this paper, we perform evolutionary simulations under such topological changes of networks and
 find that cooperation is decreased to the greatest extent
 when targeted removal and random addition of nodes are combined.
 In contrast, cooperation is maintained even at a high temptation to defect
 when random removal and preferential addition are combined.
 We also show that the degree variance, which measures the network heterogeneity,
 directly controls the robustness of cooperation.
 We find that the effect of the degree mixing pattern
 of the network is not the primary factor
 for the robustness of cooperation under
 a continuous change of network topology
 due to consequential removal and addition of nodes,
 which is quite different from the cases observed in static networks.

 This paper is organized as follows. In Sec.~\ref{sec:Model},
 we introduce a model in which removal and addition processes
 are considered on scale-free networks.
 In Sec.~\ref{sec:Result}, 
 we present the numerical results for
 the robustness of cooperation under such topological changes
 and an analysis of the results from the view point of network
 heterogeneity defined by the degree variance.
 We also investigate the effect of the degree mixing pattern
 on the evolution of cooperation.
 The summary and conclusion are given in Sec.~\ref{sec:Summary}.
 
 \section{Model}
 \label{sec:Model}
 
 To incorporate the network heterogeneity in the degree distribution
 observed in real networks, we employ the Barab\'{a}si-Albert method
 for generating initial networks
 in numerical experiments \cite{BarabasiAlbert1999}.
 Starting from a complete graph with a given small number of nodes $m_0$,
 a new node with $m \le m_0$ links is added at every time step.
 This new node is connected to $m$ existing nodes selected
 according to the probability
 $p_i = k_i /\sum k_i$,
 where $k_i$ is the degree of node $i$ of each selected node.
 Thus, nodes with a larger degree are more likely to be selected,
 hence the ``preferential attachment.''
 After $t$ discrete time steps,
 the resulting network consists of $N = t + m_0$ nodes and $mt$ links
 according to a power-law degree distribution
 with an exponent of 3 \cite{BarabasiAlbert1999}.

\begin{figure}[!t]
   \centering
   \includegraphics[width=\columnwidth]{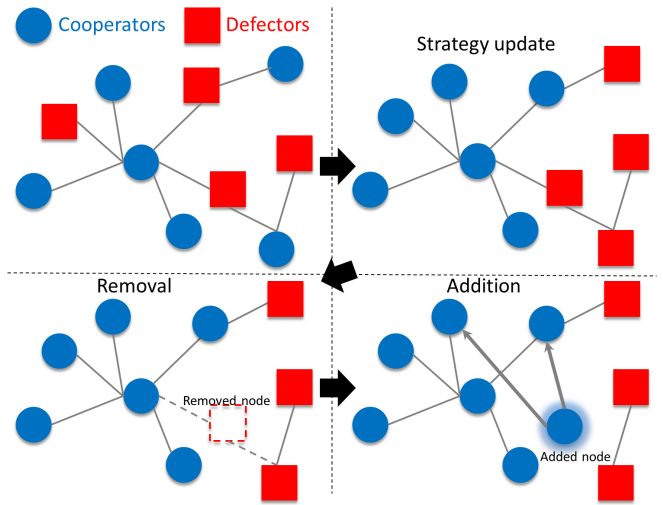}
   \caption{(Color online) An example of one ``generation'' of
   the evolutionary game considered in this paper.
   First, all individuals update their strategies simultaneously.
   Then, one defector with two links is selected for removal in this case.
   Finally, one cooperator (the strategy is randomly selected) is added
   and connected by two links to the existing nodes.
   This ``generation'' is repeated up to a given number of steps.
   We consider four models (RR, RP, TR, TP) for the removal and addition of nodes.
   }
   \label{model}
  \end{figure}

 We investigate the PD game
 on this initially scale-free network.
 Let $N$ be the total number of nodes occupied by individuals;
 each of the nodes has its strategy classified as
 either C (cooperator) or D (defector).
 Initially, both strategies C and D are randomly and equally distributed
 among the nodes of the network.
 Each node $i$ plays PD with all of its $k_i$ neighbors.
 The payoffs of the game are the following.
 Both individuals obtain $R$ for mutual cooperation and $P$ for mutual defection.
 If one selects cooperation and the other selects defection,
 the cooperator obtains $S$ as the sucker of the defector,
 and the defector obtains $T$ as the reward for temptation to defect.
 The order of the four payoffs is $T>R>P>S$ in PD.
 The sum of the payoff of individual $i$
 against its $k_i$ neighbors is denoted by $P_{i}$.
 Following Nowak and May \cite{NowakMay1992},
 we set $P=0$, $T=b>1$, $R=1$, and $S=0$,
 where $b$ is the temptation to defect.
 Next, one randomly chosen neighbor of $i$, denoted by $j$,
 also plays PD with its neighbors and obtains the payoff $P_j$.
 If $P_i < P_j$, individual $i$ imitates individual $j$'s strategy
 with probability $(P_j - P_i)/[(T-S)k_{\mathrm{max}}]$,
 where $k_{\mathrm{max}}$ is the largest degree between $i$ and $j$.
 This update principle of strategy has been adopted
 in various studies \cite{SantosPacheco2005,Rong_etal2007,Perc2009}.
 All individuals update their strategies simultaneously at each time step.
 After this update,
 the network topology is altered by one removal and one addition of nodes.
 Here we consider the following four combinations of node
 removal and addition.
 First, an existing node is removed in two different ways,
 namely, \textit{random} removal and \textit{targeted} removal.
 In the random removal, one randomly selected node is removed.
 In the targeted removal, a node with the largest degree is removed.
 In both cases, the links connected to the removed node are also removed from the network.
 After the removal, a new node is added in two different ways,
 namely, \textit{random} addition and \textit{preferential} addition.
 In the random addition, a new node connects to $m$ randomly selected existing nodes.
 In the preferential addition, a new node of $m$ degree connects to each existing node with probability $p_i = k_i/\sum_{i} k_i$.
 Because the number of removed links is preserved,
 the remaining links other than $m$ are also connected in each manner.
 We classify these four different combinations of
 node removal and addition,
 which cause continuous alteration of the network topology,
 into the following four models:
  \begin{enumerate}
   \item \textit{Random removal and Random addition (RR)}.
	 After the removal of one randomly selected node of degree $n$,
	 a new node of degree $m$ is added and connected
	 to $m$ randomly selected existing node.
	 If $m < n$,
	 each remaining $n-m$ link is connected from a randomly selected node
	 (referred to as the source) to a randomly selected node
	 (referred to as the target).
	 If $m \geq n$, only $n$ links of the added node are connected
	 to $n$ existing nodes.
	 For $n=0$, a new node immediately becomes
	 an isolated node after it is added.
	 This linking principle is also applied to the other three models.
	 
   \item \textit{Random removal and Preferential addition (RP).}
	 After the removal of one randomly selected node of degree $n$,
	 a new node of degree $m$ is added and connected to each existing node
	 with probability $p_i = k_i/\sum_{i} k_i$,
	 which is proportional to the degree of node $i$.
	 If $m < n$,
	 each remaining $n-m$ link is connected
	 from a randomly selected source node
	 to a target node with probability $p_i$.	
	 If $m \geq n$, only $n$ links of the added node are connected
	 to $n$ existing nodes.
 
   \item \textit{Targeted removal and Random addition (TR).}
	 After the removal of the node with the largest degree $n$
	 among the existing nodes,
	 a new node of degree $m$ is added and connected
	 to $m$ randomly selected existing nodes.
	 If $m < n$,
	 each remaining $n-m$ link is connected
	 from a randomly selected source node
	 to a randomly selected target node.	
	 If $m \geq n$, only $n$ links of the added node are connected
	 to $n$ existing nodes.

   \item \textit{Targeted removal and Preferential addition (TP).}
	 After the removal of the node with the largest degree $n$
	 among the existing nodes,
	 a new node of degree $m$ is added and connected to each existing node
	 with probability $p_i = k_i/\sum_{i} k_i$.
	 If $m < n$,
	 each remaining $n-m$ link is connected
	 from a randomly selected source node
	 to a target node with probability $p_i$.	
	 If $m \geq n$, only $n$ links of the added node are connected
	 to $n$ existing nodes.
  \end{enumerate}
  Note that for all four models,
  both the total number of nodes and the total number of links remain unchanged.
  In contrast, the network topologies do change.
  The strategy of a newly added node is randomly chosen
  from strategies C and D.
  The PD game of all existing nodes, updating their strategies, and
  the node removal and addition procedure make up one entire process
  in a numerical experiment, which we refer to as ``generation.''
  This generation is repeated up to a given number of steps.
  Figure~\ref{model} shows a schematic picture of one generation.
  
 \section{Results and Discussions}
 \label{sec:Result}
 
 To generate the initial networks according to the BA method,
 we took $m = m_0 = 2$ and added nodes up to $N=5000$.

\begin{figure}[!t]
  \centering
  \includegraphics[width=\columnwidth]{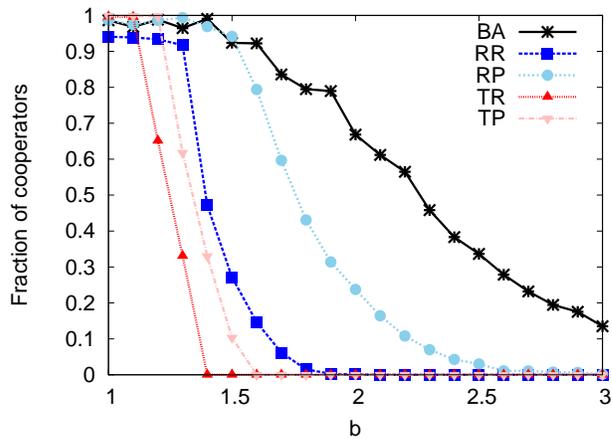}
  \caption{(Color online) Fraction of cooperators as a function of the temptation to defect $b$
  in our four different models and the original BA model averaged over 20 independent samples.
  In the BA model, the structure is kept unchanged during all generations.}
  \label{fracC}
 \end{figure}

 In Fig.~\ref{fracC},
 we plot the fraction of cooperators
 as a function of the temptation to defect $b$ for the four models.
 The results for each sample are obtained by averaging over 1000 generations
 after a transient time of 10000 generations.
 The final results are obtained by averaging over 20 independent samples
 for each set of parameters.
 The fraction of cooperators shows quite different profiles
 depending on the model.
 We also plot the case of the original BA model.
 This case always shows the highest level of cooperation
 because its hub structures, which benefit for cooperation,
 are not altered.
 The evolution of cooperation
 in models containing targeted node removal (TR and TP)
 is considerably suppressed even for small values of the temptation to defect $b$.
 The evolution of cooperation is also suppressed
 in models containing random addition of nodes (TR and RR).
 Thus, the fraction of cooperators is the most fragile in the TR model,
 which is the combination of targeted removal and random addition.
 In contrast,
 the fraction of cooperator has relatively large values
 in the RP model, which is the opposite combination of TR,
 even in the region with rather large values for the temptation to defect.

 The qualitative reason for this difference in
 the profiles of the fraction of cooperators is the following.
 It is commonly known that the network heterogeneity determines
 the fate of cooperation \cite{Perc2009}.
 If cooperators occupy the hubs of a network surrounded by other cooperators,
 their payoffs are considerably higher than other individuals.
 Cooperative hubs can therefore easily spread their strategy
 to the surrounding nodes.
 Since the fraction of hubs is extremely small even in a scale-free network,
 it is rare that a choice for random node removal hits a hub.
 Thus, cooperative hubs are maintained in random node removal.
 Moreover, the preferential addition tends to increase the degree of the hubs,
 which contributes to the resiliency of cooperation
 by expanding the network heterogeneity.
 This is the reason of the resiliency of the fraction of
 cooperators in the RP model.
 The reason of the fragility in the TR model is completely opposite
 to the case of the RP model.
 For quantitative support for this reasoning,
 we next examine the network characteristics relating
 to the network heterogeneity corresponding to the four models.

\subsection{Network heterogeneity}

The network heterogeneity is represented by the degree variance
$V = [(1/N)\sum k_i^2 - \bar{k}^2]/\bar{k}$, where $\bar{k}=(1/N)\sum k_i$.
This value becomes zero if the all nodes have the same degree
while it takes a larger value if some
nodes have an extremely large degree, such as hubs.

\begin{figure}[!t]
 \centering
 \includegraphics[width=\columnwidth]{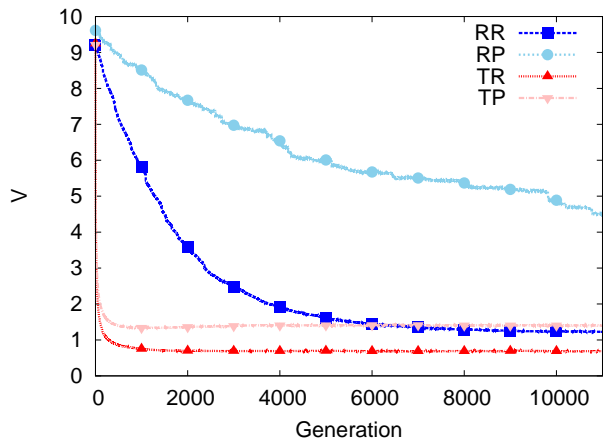}
 \caption{(Color online) The degree variance as a function of the generation
 in the four models averaged over 20 independent runs.}
 \label{DV}
\end{figure}

Figure \ref{DV} shows the degree variance as a function of generation
in the four models.
In all models, the degree variance decreases as the generation increases
from the largest degree variance of the initial BA network.

The details of the collapse of the degree variance, however,
are different in the four models.
The degree variances in TR and TP show drastic decrease
in the early stages of generation
because the largest hub is always removed in the targeted models.
On the other hand, the values of the degree variance in RR and RP
do not show such a drastic decrease.
In RP, in particular, a node is randomly removed without paying attention to its degree,
and the preferential addition of node introduces new heterogeneity.
The network heterogeneity that supports the fraction of cooperators is
thus mostly maintained in the RP model.

\begin{figure}[!t]
 \centering
 \includegraphics[width=\columnwidth]{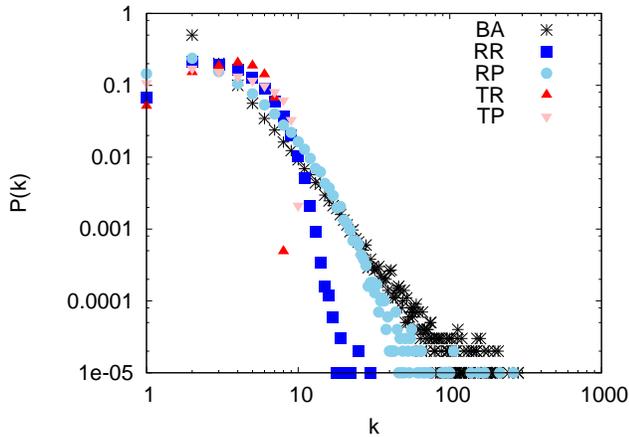}
 \caption{(Color online) The degree distribution at the final generation in our four models
 along with the degree distribution of the initial BA model
 averaged over 20 independent runs.}
 \label{DD}
\end{figure}

In Fig.~\ref{DD}, we compare the final degree distribution
of the four models to the initial BA model.
The RP model maintains some hubs, while the other three models do not.
By conducting further network analysis,
we find that the four network topologies
are completely altered from that of the original BA model (Supplemental Fig.~S1),
but the hubs in RP, which cause the network heterogeneity,
are still maintained.
This supports the result that cooperation is robust in RP \cite{Suppl}.

\subsection{Effect of the degree mixing pattern}


It has been realized that the degree correlation between the node connection
represented by the degree mixing pattern
sometimes considerably modifies the results obtained
from the mean field analysis
based only on the degree distribution \cite{Serrano_etal2006, Goltsev_etal2008, ShirakiKabashima2010, Ostilli_etal2011, Tanizawa_etal2012}.
In static networks, it is known that an uncorrelated network promotes cooperation \cite{Rong_etal2007}.
Here we investigate the effects of degree correlation
on the resilience of the clusters of cooperators
in the present cases in which the network topologies
are constantly changed.

According to Newman,
we measure the degree correlation of a network by
the Pearson coefficient $r_k$ \cite{Newman2002}. 
If $r_k$ is positive,
nodes with almost the same degree tend to be connected;
the correlation is denoted as ``assortative.''
In assortative networks, hubs tend to be connected to other hubs.
If $r_k$ is negative,
nodes with different degrees tend to be connected;
the correlation is denoted as ``disassortative.''
In disassortative networks, hubs tend to be connected to nodes
with small degrees.
Newman pointed out that
the Pearson coefficient of the BA model takes a very small value,
$r_k \approx 0$,
which means that the BA networks are almost uncorrelated \cite{Newman2002}.

\begin{figure}[!t]
 \centering
 \includegraphics[width=\columnwidth]{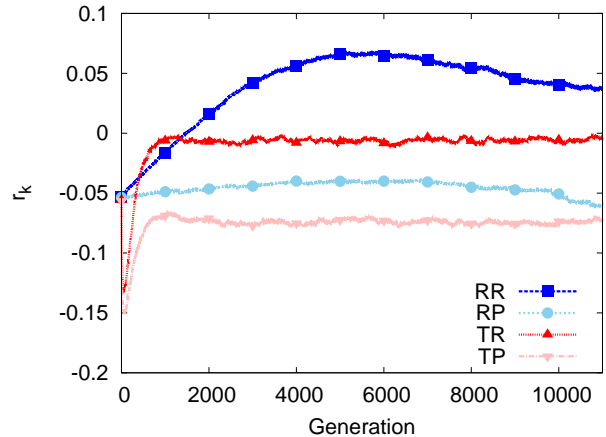}
 \caption{(Color online) The correlation coefficient ($r_k$) as a function of the generation
 in the four models averaged over 20 independent runs.}
 \label{DC}
\end{figure}

Figure \ref{DC} shows the variation of the correlation coefficient
with respect to the generation.
We see three different correlation regimes:
disassortative (RP and TP), uncorrelated (TR), and assortative (RR).
By reexamining the results for the fraction of cooperators (Fig.~\ref{fracC})
in terms of degree correlation (Fig.~\ref{DC}),
the RP model, in which the fraction of cooperators is most robust,
falls in the disassortative regime.
It does not seem, however, that the degree correlation plays a key role
in the robustness of the cooperation
since the fraction of cooperators in the TP model,
which also falls in the disassortative regime, is rather fragile.
For the robustness of cooperation,
the resiliency of the hubs with the largest degree controlling
the stability of cooperation is most important.
In this regard, elimination of the hubs due to targeted attack
is most fatal to the robustness of the cooperation.
On the contrary, random node removal
rarely hits the hubs in elimination.
This is the main reason for
the difference between RP and TP in the PD game on networks
with continuously changing of network topology.
The fate of cooperation is thus dominated by the network heterogeneity
and the degree correlation seems to be
a secondary factor in the dynamic network.

It should be noted that the fraction of cooperators is most fragile
in the TR model, which falls in the uncorrelated regime.
This result is different from the analysis of static networks,
where uncorrelated networks have an advantage in terms of the robustness
of the clusters of cooperators \cite{Rong_etal2007}.

 \section{Summary}
 \label{sec:Summary}
 
 The evolution of cooperation is still an open question in various fields.
 It is commonly accepted that the network structure is
 one of the main controlling factors for the promotion of cooperation.
 In social or ecological systems in the real world,
 it is plausible to assume that an individual
 or a species represented by a node is constantly
 replaced by or added to another,
 which introduces a continuous topological
 change in the network structure.
 It is therefore important to know whether cooperation is maintained
 under such circumstances.
 
 Based on these motivations,
 we numerically investigated the robustness of cooperation on scale-free networks
 under continuous topological changes
 due to the removal and addition of nodes in a network.
 We have found that cooperation is most robust against
 random removal and preferential addition of nodes,
 while cooperation is most vulnerable against targeted attack.
 The damage caused by the targeted attack is not fully compensated
 by either random or preferential addition. 
 By calculating several network characteristics,
 we have revealed that the network heterogeneity dominates
 the fate of cooperation.
 If the degree variance is large, cooperation is maintained.
 We have also shown that the degree correlation
 does not affect the cooperation much on dynamical networks because
 cooperation mainly depends on the existence of cooperative hubs,
 which shows a sharp distinction from the cases observed in static networks.
 These results might explain the fact that a vast variety of individuals is needed in a society
 where many individuals independently join and leave because hubs are actually important for maintaining cooperation on
 an online friendship network \cite{Fu_etal2007}.

 \section*{Acknowledgment}
 
 T.T., acknowledges
 the support of a Grant-in-Aid for Scientific Research (C)
 (Grant No.\ 24540419) from the Japan Society for the Promotion of Science.

\providecommand{\noopsort}[1]{}\providecommand{\singleletter}[1]{#1}%

\end{document}